\documentclass[aps,prb,showpacs,twoside,amsmath,amssymb,floatfix,superscriptaddress,twocolumn,reprint]{revtex4-1}
\usepackage{graphicx}
\usepackage{txfonts}
\usepackage{bm}
\DeclareMathOperator{\sgn}{sgn}
\DeclareMathOperator{\dotproduct}{\cdot}
\DeclareMathOperator{\Imag}{Im}
\usepackage{units}

\usepackage[bookmarks=true,colorlinks=true,urlcolor=blue,linkcolor=blue,citecolor=blue,breaklinks]{hyperref} 
\usepackage[usenames,dvipsnames]{color}
\usepackage{bbold}
\begin{document}
\newcommand{\brho}{{\bm \rho}}
\def \beps{{\hbox{\boldmath $\epsilon$}}}
\def \bdelta{{\hbox{\boldmath $\delta$}}}
\pacs{73.22.Pr, 73.22.Lp}

\title{Using surface plasmonics to turn on fullerene's dark excitons}
\author{V. Despoja}
\email{vito@phy.hr}
\affiliation{Department of Physics, University of Zagreb, Bijeni\v{c}ka 32, HR-10000 Zagreb, Croatia}
\affiliation{Donostia International Physics Center (DIPC), Paseo de Manuel de Lardizabal 4, ES-20018 San Sebast{\'{i}}an, Spain}
\author{D. J. Mowbray}
\email{duncan.mowbray@gmail.com}
\affiliation{Donostia International Physics Center (DIPC), Paseo de Manuel de Lardizabal 4, ES-20018 San Sebast{\'{i}}an, Spain}
\affiliation{Nano-Bio Spectroscopy Group and ETSF Scientific Development Center, Departamento de F{\'{\i}}sica de Materiales, Universidad del 
Pa{\'{\i}}s Vasco UPV/EHU, ES-20018 San Sebasti{\'{a}}n, Spain}
\begin{abstract}
Using our recently proposed Bethe-Salpeter $G_0W_0$ formulation, we explore the optical absorption spectra of fullerene (C$_{60}$) near coinage metal surfaces (Cu, Ag, and Au). 
We pay special attention to how the surface plasmon $\omega_S$ influences the optical activity of fullerene.  We find the lower energy fullerene excitons at $3.77$ and $4.8$~eV only weakly interact with the surface plasmon. However, we find the surface plasmon strongly interacts with the most intense fullerene $\pi$ exciton, i.e.\ the dipolar mode at $\hbar\omega_+\approx6.5$~eV, and the quadrupolar mode at  $\hbar\omega_-\approx6.8$~eV.  When fullerene is close to a copper surface ($z_0\approx5.3$~\AA) the dipolar mode $\omega_+$ and ``localized'' surface plasmons in the molecule/surface interface hybridize to form two coupled modes which both absorb light. As a result, the molecule gains an additional optically active mode. Moreover, in resonance, when $\omega_S\approx\omega_\pm$, the 
strong interaction with the surface plasmon destroys the $\omega_-$ quadrupolar character and it becomes an optically active mode. In this case the molecule gains two additional very intense optically active modes. Further, we find this resonance condition, $\omega_S \approx \omega_\pm$, is satisfied by silver and gold  metal surfaces
.  
\end{abstract}

\maketitle

\section{Introduction}

Using molecules to absorb light within a photovoltaic device allows the use of chemical functionalization to engineer the energy of light being absorbed\cite{PolymerSolarCellsRev,DyeSensitizedSolarCellsRev}.  The goal is to have several different types of molecules within the same device, each absorbing at different wavelengths, to cover the entire visible spectra.  However, unlike solid-state based devices, the separation of a generated electron-hole pair, or exciton, is a key obstacle for a molecular absorber.  A functioning device needs to reduce the recombination rate, or gemination, of the molecule's exciton.  This is accomplished through transfer of the electron (or hole) to a substrate or acceptor molecule.  

An exhaustive experimental search of the vast parameter space of possible molecular absorbers and charge acceptors has proved daunting.  Such a problem is ideally suited for a computational screening approach.  However, standard computational methods, such as density functional theory (DFT), have difficulties describing the optical absorption levels of molecules.  On the other hand, more advanced methods, such as solving the Bethe-Salpeter equation (BSE) using quasiparticle $G_0W_0$ eigenvalues \cite{Hedin,Strinati,GWtheory,Rubio,StevenLouie98,StevenLouie2000}, are computationally unfeasible for screening studies.  In fact, a BSE-$G_0W_0$ treatment is necessary to describe the complex renormalization of molecular energy levels due to the anisotropic screening at an interface \cite{StevenLouie2006,JuanmaRenormalization1,JuanmaRenormalization2,GiustinoPRL2012,OurJACS,MiganiLong,MiganiInvited}.

But when the molecule and substrate are well separated, the inter-system electronic overlap may be neglected.  By describing the substrate via its response function and the molecule within BSE-$G_0W_0$, one may obtain an accurate description of the molecule/surface interactions at a reasonable computational cost \cite{Spataru,Exciton}. Within such a reformulation of BSE-$G_0W_0$ for weak molecule-substrate coupling, it is now possible to do computational screening of optical absorption and interfacial coupling of a molecule near a surface.

In fact, it is precisely this weak coupling regime which is of interest for nanoplasmonic single-molecule sensing\cite{biosensors,LabelFreePlasmonicBiosensors2009,LSPRS_ChemRev,SingleMoleculeDetection2012}.  For localized surface plasmon resonance sensors (LSPRS), it is now possible to detect single molecules which are not even adsorbed on the substrate.  However, to design LSPRs which can not only differentiate between single molecules, but also their height off the surface, would require a computational screening approach.  Within our BSE-$G_0W_0$ reformulation, we are now able to find the properties the substrate needs in order to have a strong interfacial coupling.  In other words, the particular type of nanoparticle, corner, step, metal alloy, etc.\ needed to have a strong hybridization with a particular molecule's excitonic levels. 

Recently, highly efficient polymer-fullerene-based organic photovoltaic devices have been demonstrated, which employ fullerenes\cite{SmalleyC60,DresselhausC60} as electron acceptors \cite{PolymerFullerenePhotovoltaic,Fullerene_apl1,Fullerene_apl2}.  Moreover, photovoltaic devices have also been recently demonstrated which are based solely on fullerenes \cite{FullereneOnlyPhotovoltaic}.  In this case, different fullerene morphologies act as optical absorbers, charge acceptors, and charge donors.  This high degree of versatility in fullerene's functionality within photovoltaic devices makes a thorough understanding of its optoelectric properties an essential test case for understanding photovoltaic processes.  

Although the adsorption of fullerene on metal surfaces has been intensively studied experimentally \cite{FullereneAg(111)_1,FullereneAg(111)_2,FullereneAl(111)}, there remains a lack of experimental studies dealing with the influence of the metal surface on the molecule's optical absorption spectra. In this work, we demonstrate how the optical absorption spectra of fullerene couples to the surface plasmon modes of a coinage metal (Cu, Ag, Au) surface.  We find the optical spectra exhibit a strong dependence on the molecule's height.  This suggests a nanoplasmonic chemical detector of Ag or Au could readily differentiate not only the presence of a single C$_{60}$ molecule, but also its height above the metal surface.  We also find the hybridization of the fullerene excitonic levels with the surface plasmons can result in not only a redistribution of their energies, but ``turn on'' the fullerene dark excitons,  essentially making them bright.

In Sec.~\ref{Optabsspe} we briefly present the theoretical methodology we use.  We first show how we obtain the 4-point polarizability matrix $L^{kl}_{ij}(\omega)$ in Sec.~\ref{optabs}. We then show in Sec.~\ref{calcoptspectra} how $L^{kl}_{ij}(\omega)$ may be used to calculate the optical absorption spectra.  A point polarizable dipole model is then developed in Sec.~\ref{pointdipole}.  In Sec.~\ref{optabsonsub} we show how these models may be applied to the optical absorption of a fullerene molecule near a metal surface. Finally, in Sec.~\ref{ComputationalDetails}, we provide details of the computations we perform.

In Sec.~\ref{Results} we present the results for the HOMO--LUMO gap and molecular optical absorption spectra as a function of the molecule's height above a copper, silver, or gold surface. This is followed by concluding remarks in Sec.~\ref{Conclusions}.

\section{Theoretical Methodology}
\label{Optabsspe}
\subsection{Solving the BSE for 4-point polarizability matrix }
\label{optabs}
When a molecule absorbs light, an electron-hole pair may be created.  In the lowest order approximation, the electron and hole can be consider as two independent particles, which without any interactions, simply propagate throughout the molecule.  Such long lived electron-hole pair propagation can be described as a convolution of two one-particle Green's functions 
\begin{equation}
L_0({\bf r}_1,{\bf r}_2;{\bf r}'_1,{\bf r}'_2,\omega)=
-i\int^{\infty}_{-\infty}\!\!\frac{d\omega'}{2\pi} G_0({\bf r}_2,{\bf r}'_1,\omega')
G_0({\bf r}_1,{\bf r}'_2,\omega+\omega').
\label{freefor}
\end{equation}
In the independent electron approximation, the Green's functions are given by
\begin{equation}
G_0({\bf r},{\bf r}',\omega)=\sum_i\frac{\psi_i({\bf r})\psi^*_i({\bf r}')}{\omega-\varepsilon_i+i\eta\sgn(\varepsilon_F-\varepsilon_i)},
\label{KSgreen}
\end{equation}
where $\psi_i({\bf r})$ and $\varepsilon_i$ are the molecular orbitals and energy levels, respectively. These may be easily calculated 
at the Kohn-Sham (KS) 
level.

Note that the excited electron and hole can still interact with other molecular excitations, e.g.\ collective electronic modes (plasmons) or molecular vibrational modes (phonons). Such additional interactions, especially long range electron-electron correlations, are not included at the KS level. The one particle Green's function of Eq.~(\ref{KSgreen}) must then be corrected in order to include 
all these effects. Moreover, because of the electron-electron interaction, the excited electron and hole can interact mutually or annihilate and interact with other electron-hole excitations in the molecule. In order to obtain an accurate molecular excitation spectra, all these processes should be carefully taken into account. 

To do so, we calculate the full electron-hole propagator or 4-point polarizability 
\begin{equation}
L({\bf r}_1,{\bf r}_2;{\bf r}'_1,{\bf r}'_2,\omega)=\sum_{ijkl}\Theta_{ij}^{kl}
L^{kl}_{ij}(\omega)\psi_i({\bf r}_1)\psi^*_j({\bf r}'_1)\psi_l({\bf r}_2)\psi^*_k({\bf r}'_2),
\label{Lexpanz}
\end{equation}
where the 4-point polarizability matrix $L^{kl}_{ij}(\omega)$ 
satisfies the Bethe-Salpeter equation \cite{Exciton,Hedin,Strinati,GWtheory,Rubio} 
\begin{equation}
L^{kl}_{ij}(\omega)=\tilde{L}^{kl}_{ij}(\omega)+
\sum_{i_1j_1k_1l_1}\Theta_{i_1 j_1}^{k_1 l_1}\ \tilde{L}^{i_1j_1}_{ij}(\omega)\ \Xi^{k_1l_1}_{i_1j_1}\ L^{kl}_{k_1l_1}(\omega).
\label{mateqforBSE}
\end{equation}
The prefactor 
\begin{equation}
\Theta_{ij}^{kl} \equiv |f_j-f_i||f_l-f_k|,
\end{equation}
ensures that only transitions between empty and filled 
molecular states contribute to $L$, where
\begin{equation}
f_i=\left\{\begin{array}{cc}1;&i\le N\\0;&i>N\end{array}\right.
\end{equation} 
is the occupation factor. The matrix of noninteracting quasiparticle 4-point polarizability has the form  
\begin{equation}
\tilde{L}^{kl}_{ij}(\omega)=
2\frac{f_j-f_i}{\omega+\tilde{\varepsilon}_j-\tilde{\varepsilon}_i+i\eta\sgn(\tilde{\varepsilon}_i-\tilde{\varepsilon}_j)}\delta_{ik}\delta_{jl},
\label{iqppol}
\end{equation}
where the factor of two is introduced to include contribution from both spin channels and $N$ is the number of occupied orbitals. The quasiparticle energies $\tilde{\varepsilon}_i$ are obtained by solving the Dyson equation\cite{Exciton},
where the exchange and correlation self-energy operator is calculated within the $G_0W_0$ approximation\cite{GWtheory}, i.e.\ 
\begin{equation}
\Sigma_{\textit{XC}}({\bf r},{\bf r}',\omega)=i\int^{\infty}_{-\infty}\frac{d\omega'}{2\pi}e^{-i\omega'\delta} 
G_0({\bf r},{\bf r}',\omega-\omega')W({\bf r},{\bf r}',\omega').
\label{GWapprox}
\end{equation} 
The BSE kernel consists of two terms 
\begin{equation}
\Xi^{kl}_{ij}=\Xi^{kl,H}_{ij}-\Xi^{kl,F}_{ij}.
\label{barculkul}
\end{equation}
The first term is the BSE-Hartree kernel, given by 
\begin{equation}
\Xi^{kl,H}_{ij}=\int d{\bf r}_1d{\bf r}_2 \phi^j_i({\bf r}_1)V({\bf r}_1-{\bf r}_2)\phi^k_l({\bf r}_2),
\label{BSEH}
\end{equation}
where $V({\bf r}_1-{\bf r}_2)$ is the propagator of the bare Coulomb interaction
and 
\begin{equation}
\phi^j_i({\bf r})=\psi^*_i({\bf r})\psi_j({\bf r}).
\label{pairfun}
\end{equation}
represents the two-particle wave functions. 
The second term is the BSE-Fock kernel, given by 
\begin{equation}
\Xi^{kl,F}_{ij}=\frac{1}{2}\int d{\bf r}_1d{\bf r}_2 \phi^j_l({\bf r}_1)W({\bf r}_1,{\bf r}_2,\omega=0)\phi^k_i({\bf r}_2).
\label{Fockker}
\end{equation}
The propagator of the dynamically screened Coulomb interaction $W({\bf r}_1,{\bf r}_2,\omega)$, which enters in Eq.~(\ref{GWapprox}) and the Fock kernel (\ref{Fockker}), is given by     
\begin{eqnarray}
W({\bf r},{\bf r}',\omega)&=&V({\bf r},{\bf r}')+
\sum_{\alpha\beta\gamma\delta}\Theta_{\alpha\beta}^{\gamma\delta}
L^{\gamma\delta,\mathrm{RPA}}_{\alpha\beta}(\omega)\nonumber\times\\
&&\int d{\bf r}_1d{\bf r}_2
V({\bf r},{\bf r}_1)\phi^\alpha_\beta ({\bf r}_1)
\phi^\delta_\gamma({\bf r}_2)
V({\bf r}_2,{\bf r}'),
\label{Wexpanz}
\end{eqnarray}
where $L^{kl,\mathrm{RPA}}_{ij}(\omega)$ is the solution within the random phase approximation (RPA).  This is equivalent to Eq.(\ref{mateqforBSE}), where we put 
$\Xi^{kl}_{ij}=\Xi^{kl,H}_{ij}$ and $\tilde{\varepsilon}_i=\varepsilon_i$.

After solving the matrix equation (\ref{mateqforBSE}), we obtain the 4-point polarizability matrix $L^{kl}_{ij}(\omega)$.  From $L^{kl}_{ij}(\omega)$ we may then calculate the optical absorption spectra.

\subsection{Calculation of the optical absorption spectra}\label{calcoptspectra}
In an optical absorption experiment the incident electromagnetic wave couples to the electronic excitations in the system and is partially absorbed. In linear response theory the power at which the external electromagnetic energy is absorbed by the system can be obtained from
\begin{equation}
P(t)=\int^{\infty}_{-\infty}dt_1\int d{\bf r}_1d{\bf r}_2\ {\bf E}^{\textit{ext}}({\bf r}_1,t)\Pi({\bf r}_1,{\bf r}_2,t-t_1){\bf A}^{\textit{ext}}({\bf r}_2,t_1),
\label{powerloss}
\end{equation}
where $\Pi$ is the current--current response function of the system, while ${\bf E}^{\mathit{ext}}$ and ${\bf A}^{\mathit{ext}}$ are the external electric field and vector potential, respectively. 

We shall assume that the incident electromagnetic field is a plane wave of unit amplitude
\begin{equation}
{\bf A}^{\textit{ext}}({\bf r},t)={\bf e}\cos({\bf k}\cdot{\bf r}-\omega t), 
\label{incidentemp}
\end{equation}
where ${\bf e}$ is the polarization vector. If we also assume there is no external scalar potential, i.e.\ $\Phi^{\textit{ext}}=0$, then from Maxwell's equations ${\bf E}^{\textit{ext}}=-\frac{1}{c}\frac{\partial{\bf A}^{\textit{ext}}}{\partial t}$. If the wavelength ${\lambda}$ is much larger than the dimension of the illuminated system or the crystal unit cell, the dipole approximation may be applied.  In this case the absorption power becomes
\begin{equation}
P(\omega)=-\omega \Imag\left\{\sum_{\mu\nu}e_\mu e_\nu\int d{\bf r}_1{\bf r}_2\Pi_{\mu\nu}({\bf r}_1,{\bf r}_2,\omega)\right\}. 
\label{apspowerm}
\end{equation}

In the Coulomb gauge ($\nabla\dotproduct {\bf A}=0$), there is an instantaneous interaction mediated by the Coulomb interaction $V$ and a transverse interaction that is retarded and mediated by photons. In small systems such as a molecule, the interaction between charge/current fluctuations mediated by photons is negligible compared to the Coulomb interaction. This allows us to describe all interactions inside the molecule by the instantaneous Coulomb interaction $V$ and the interaction of the molecule with the environment by both interactions.  In this case interactions are only with photons described by ${\bf A}^{\textit{ext}}$.

As a result, the current--current response function can be expressed in terms of the 4-point polarizability matrix  
\begin{equation}
\Pi_{\mu\nu}({\bf r},{\bf r}',\omega)=
\frac{e^2\hslash}{m^2c}\sum_{ijkl}\Theta_{ij}^{kl}L^{kl}_{ij}(\omega)\psi^*_j({\bf r})\nabla_\mu\psi_i({\bf r})\psi^*_k({\bf r}')\nabla_\nu\psi_l({\bf r}').
\label{Piexpanz}
\end{equation}

After inserting (\ref{Piexpanz}) into (\ref{apspowerm}), the absorption power becomes: 
\begin{equation}
P(\omega)=-\omega \Imag\left\{\sum_{ijkl}\Theta_{ij}^{kl}L^{kl}_{ij}(\omega)J_{ji}J_{kl}\right\},
\label{apspowerFm}
\end{equation}
where the form factors are
\begin{equation}
J_{ij}=\sum_{\mu}e_\mu\int d{\bf r} \psi^*_i({\bf r})\nabla_\mu\psi_j({\bf r}).
\label{Formfact}
\end{equation}

\subsection{Point polarizable dipole model}
\label{pointdipole}

We shall next develop a simple model for the case where there is strong spatial localization of the interacting plasmas in the molecule ($M$) and surface ($S$), compared to the wavelength of the incident light. In this case we may use a simple model for the optical absorption based on point 
polarizable dipoles. Here the molecule is modeled by a point polarizable dipole of frequency $\omega_M$ and the adjacent surface plasma is modelled by another point polarizable dipole of frequency $\omega_S$.  Their density--density response functions may then be written as   
\begin{equation}
\chi_i(\omega)=\frac{2\omega_i}{\omega(\omega+i\eta)-\omega^2_i},\ \ i=M,S.
\label{dipol1}
\end{equation}    
where $M$ denotes the molecular and $S$ denotes the surface response functions, respectively.
If we suppose the dipoles are mutually interacting, then the molecular response function $\chi_M$ is renormalized by the surface.  In this case the renormalized molecular density--density response function may be expressed as \cite{SSC,JPCM}
\begin{equation}
\tilde{\chi}_{M}(\omega)=\frac{\chi_M(\omega)}{1-V^2\chi_M(\omega)\chi_S(\omega)}
\label{dipol2}
\end{equation}
where $V$ represents the strength of the dipole--dipole interaction.
In analogy with Eqs.~(\ref{Piexpanz}--\ref{Formfact}), the molecular current--current response function 
is then $\tilde{\Pi}(\omega)=J^2\tilde{\chi}_{M}(\omega)$, where $J$ represents the current vertice\cite{Kupcic}. 
The molecular absorption spectra then becomes 
\begin{equation}
P(\omega)=-\omega \tilde{\Pi}(\omega).
\label{dipol3}
\end{equation}

\subsection{Optical absorption spectra of fullerene near a metal surface} 
\label{optabsonsub}

\begin{figure}
\centering
\includegraphics[width=0.8\columnwidth]{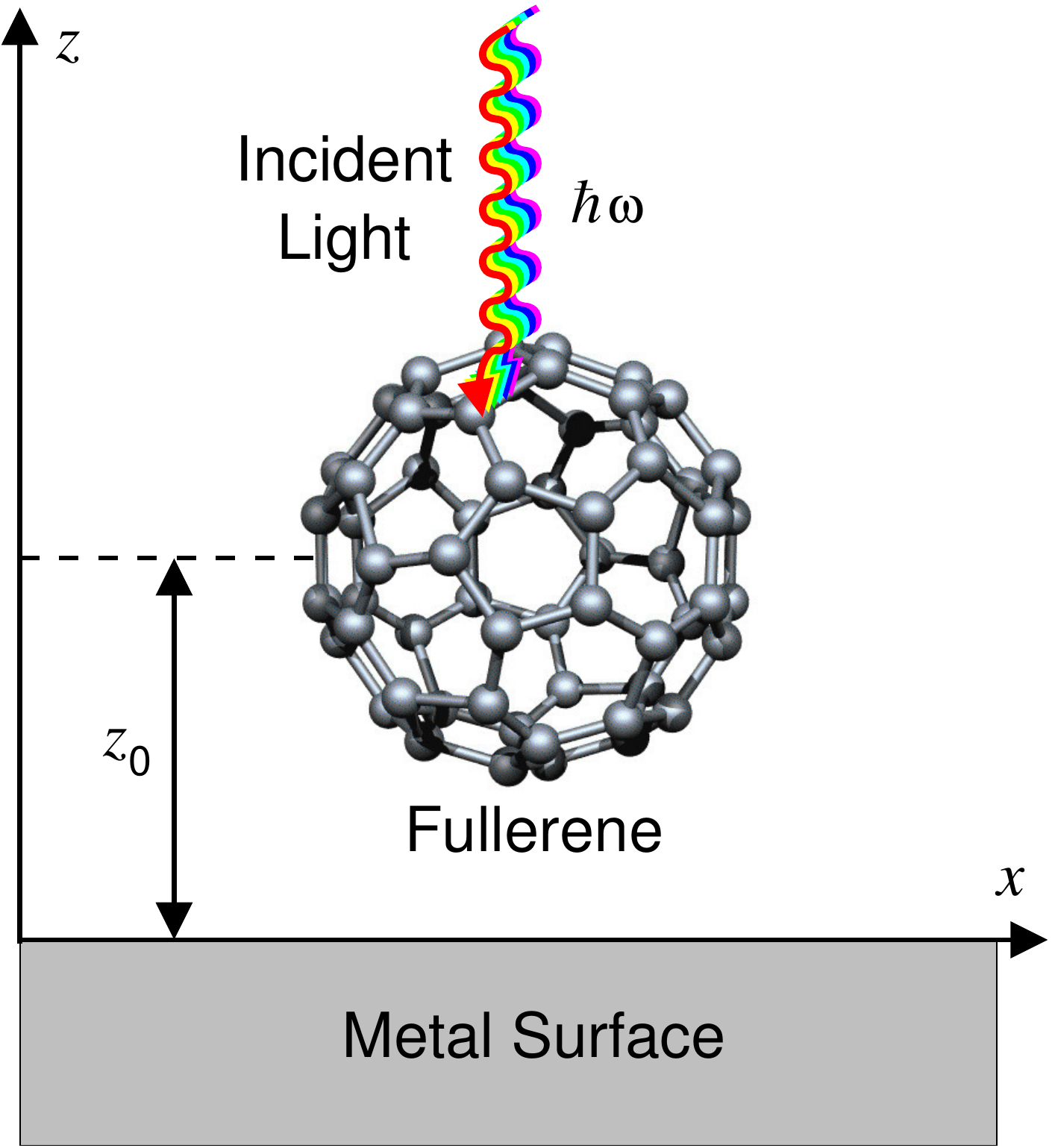}
\caption{(Color online) Schematic depicting $x$-polarized incident light of energy $\hbar\omega$ being absorbed by a fullerene molecule at a height $z_0$ above a metal surface.}
\label{Fig1}
\end{figure}

We next investigate the optical absorption of a fullerene molecule near a metal surface.  This process is depicted schematically in Fig.~\ref{Fig1}. 
The incident electromagnetic field may induce currents in the molecule.  If the molecule is placed sufficiently close to a conducting surface, these currents can induce charge density fluctuations in the surface. The induced surface charge produces an electromagnetic field in response.  This field screens the interaction between the charge density fluctuations in the molecule, and causes a renormalization of the molecular optical and quasiparticle spectra. Here we shall briefly explain the modifications which should be done within the BSE-$G_0W_0$ scheme in order to properly include the polarization of the metallic surface. 

The plane of the metal surface is aligned with the $xy$-plane, i.e.\ has a normal parallel to the $z$-axis, as depicted in Fig.~\ref{Fig1}. The ground state electronic structure of the surface is treated in a jellium model \cite{Leo}, where the jellium edge is located at 
$z=0$. The height of the fullerene $z_0$ is defined as the distance between the fullerene's centroid and the jellium edge of the metal surface, as shown in Fig.~\ref{Fig1}. 

Here we consider the case of a molecule $z_{0}\gtrsim5$~\AA\ above the metal surface. This means the electronic densities of the molecule and the surface  do not overlap. This fact simplifies the impact of the metal surface on the 4-point polarizability matrix calculation significantly. 
Since there is no inter-system electron hopping, the molecule and surface can be treated as two separated systems.  In this case, the molecule and surface only interact via the long range Coulomb term, through which there can be mutual polarization. All interactions that propagate inside the molecule can thus be additionally screened by polarization of the surface. This means that the 
bare Coulomb interaction that propagates inside the molecule should be renormalized by the surface screening, i.e.\ 
\begin{equation}
V({\bf r},{\bf r}')\rightarrow\tilde{W}({\bf r},{\bf r}',\omega)=V({\bf r},{\bf r}',\omega)+\Delta W({\bf r},{\bf r}',\omega),
\label{indc}
\end{equation}
where $\Delta W$ represents the induced Coulomb interaction of the metallic surface \cite{Exciton}. 

The induced Coulomb interaction $\Delta{W}$ can be Fourier transformed in the $xy$-plane   
\begin{equation}
\Delta{W}({\bf r},{\bf r}',\omega)=\int\frac{d{\bf Q}}{(2\pi)^2}e^{i{\bf Q}(\brho-\brho')}\Delta{W}(\textbf{Q},\omega,z,z'),
\label{propex}
\end{equation}
where $\brho=(x,y)$ and ${\bf Q}=(Q_x,Q_y)$ is a two-dimensional wave vector. 
In the region $z,z'>0$, the molecule feels an ``external'' field from the metal for which the spatial part of the Fourier transform (\ref{propex}) has the simple 
form \cite{Leo,Physcri}
\begin{equation}
\Delta{W}({\bf Q},\omega,z,z')=D(Q,\omega)e^{-Q(z+z')}.
\label{ukr0}
\end{equation}
The surface excitation propagator $D(Q,\omega)$ contains the intensities of all (collective and single particle) electronic excitations in the metal surface. The details of the calculation of 
the propagator $D(Q,\omega)$ can be found in Ref.~\onlinecite{Physcri}. The two-dimensional ${\bf Q}$ integration in (\ref{propex}) is performed using a $61\times61$ rectangular mesh and the cutoff wave vector $Q_C\approx0.57$~\AA$^{-1}$. 

In this way, the renormalized 4-point polarizability matrix $\tilde{L}^{kl}_{ij}(\omega)$ is obtained by solving the same BSE matrix equation (\ref{mateqforBSE}).  
However, the bare Coulomb interaction which enters in Hartree and Fock kernels (\ref{BSEH}) and (\ref{Fockker}) is corrected by 
the induced Coulomb interaction $\Delta W({\bf r},{\bf r}',\omega)$. 
The molecular absorption spectra is then obtained by using (\ref{apspowerFm}) in which enters the renormalized 4-point polarizability matrix 
$\widetilde{L}^{kl}_{ij}(\omega)$. This means that in this model the incident electromagnetic field drives only the molecule directly, although the surface modes 
are driven indirectly via excitation of the molecule.

In this model, we exclude the direct interaction of the incident electromagnetic waves with the surface. The direct influence of the metal surface on the molecular absorption spectra can be understood in terms of a simple Drude model. In this case, the in-plane ($x$ or $y$) polarized light should be completely reflected or transmitted, if we neglect ohmic losses. Incident light at frequencies below the plasmon frequency $\omega_p$ of the metal surface ($\omega<\omega_P$) is completely reflected, while for $\omega>\omega_P$ it is completely transmitted. Therefore, the direct influence of the metal surface on molecular absorption in the region $\omega<\omega_P$ is simply an enhancement by a factor of two, while for $\omega>\omega_P$ the light passes through the metal and does not affect the molecular absorption. As we are primarily concerned with the region below $\omega_p$ ($\hbar\omega_p \approx 10.6$~eV for copper and $\hbar\omega_p \approx 9.1$~eV for gold and silver \cite{AshcroftMermin}), the direct influence of the metal surface on the molecular absorption is simply a rescaling of the intensity by two. This means that the direct interaction between the incident light and the metal surface does not affect the molecule/surface absorption spectrum qualitatively, and may be neglected. 

Interactions with the surface also renormalize the quasiparticle energy levels 
$\tilde{\varepsilon}_i$. In the lowest order approximation, this can be done in a such way that the self energy operator (\ref{GWapprox}) is corrected by the induced self energy operator  
\begin{equation}
\Delta\Sigma_{\textit{XC}}({\bf r},{\bf r}',\omega)=i\int^{\infty}_{-\infty}\frac{d\omega'}{2\pi}e^{-i\omega'\delta} 
\widetilde{G}_0({\bf r},{\bf r}',\omega-\omega')\Delta W({\bf r},{\bf r}',\omega').
\label{indGWapprox}
\end{equation}
This implies that the induced self energy of $i^{\mathrm{th}}$ state becomes      
\begin{equation}
\Delta\Sigma^{\textit{XC}}_{i}(\omega)=\Delta\Sigma_i^X(\omega)+\Delta\Sigma^{C}_i(\omega),
\label{indxc}
\end{equation}
where the induced exchange self energy becomes    
\begin{equation}
\Delta\Sigma_i^X(\omega)=-\sum^{N}_{j=1}\Delta W^{ij}_{ij}(\omega-\tilde{\varepsilon}_j)
\label{graingx}
\end{equation}
and the induced correlation term is  
\begin{eqnarray}
\Delta\Sigma^{C}_i(\omega)=-\frac{1}{\pi}\sum^{\infty}_{j=1}\int^{\infty}_0d\omega' \frac{\Imag\left\{\Delta W^{ij}_{ij}(\omega')\right\}}{\omega-\tilde{\varepsilon}_j-\omega'+i\eta}.
\label{indcorr}
\end{eqnarray}
The induced Coulomb interaction matrix elements are then
\begin{equation}
\Delta{W}^{kl}_{ij}(\omega)=\int_{\Omega_{\text{cell}}}d{\bf r}_1d{\bf r}_2 \phi^j_i({\bf r}_1)\Delta{W}({\bf r}_1,{\bf r}_2,\omega)\phi^k_l({\bf r}_2).
\label{indmatrel}
\end{equation}
Note that the Green's function appearing in (\ref{indGWapprox}) is the renormalized Green's function $\widetilde{G}_0$ in which enters the 
quasiparticle $G_0W_0$ eigenenergies $\tilde{\varepsilon}_i$ obtained for the isolated molecule. Accordingly, in (\ref{graingx}) and 
(\ref{indcorr}) the renormalized quasiparticle energy levels $\tilde{\varepsilon}_j$ also appear. 

\subsection{Computational Details}\label{ComputationalDetails}

Calculations of the isolated fullerene molecule have been performed with the DFT code \textsc{vasp}\cite{kresse1996b}  within the projector augmented wave (PAW) scheme \cite{kresse1999}, using the local density approximation (LDA) \cite{LDA} for the exchange and correlation (xc)-functional.  We model the molecule using a periodically repeated $24.18\ \AA \times 24.18\ \AA\times 24.18\ \AA$ unit cell. Since there is no intermolecular overlap, the ground state electronic density is calculated at the $\Gamma$ point only.  The geometries have been fully relaxed, with all forces $\lesssim$ 0.02 eV/\AA.  We employ a plane-wave energy cutoff of 445 eV, an electronic temperature $k_B T\approx0.2$ eV with all energies extrapolated to $T\rightarrow 0$ K, and a PAW LDA pseudopotential for carbon.  

To calculate the quasiparticle $G_0W_0$ eigenvalues $\tilde{\varepsilon_i}$ for the isolated fullerene molecule, one must include an increased number of unoccupied states to describe the continuum.  The fullerene molecule has $240$ valence electrons, which corresponds to $120$ doubly occupied valence orbitals. We found the inclusion of 576 bands, i.e.\ 7.6 unoccupied bands per atom, provided converged values for $\tilde{\varepsilon_i}$.  The screening $W$ is obtained from  the  dielectric function, based on the KS wavefunctions \cite{Angel}.  This is calculated using linear response time-dependent DFT within RPA, including local field effects \cite{KresseG0W0}.  To calculate the dielectric function\cite{KresseG0W0} we employed an energy cutoff of 40 eV for the number of \textbf{G}-vectors, and a non-linear sampling of 40 frequency points for the dielectric function up to 200 eV.  This large energy range is necessary to include the main features of fullerene's dielectric response \cite{Scully05,Moskalenko2012}.  From these calculations we obtained converged quasiparticle $G_0W_0$ eigenvalues $\tilde{\varepsilon}_i$ for the isolated fullerene molecule.

The fullerene KS orbitals $\psi_i({\bf r})$ are obtained by using the plane-wave self-consistent field DFT code \textsc{PWscf} of the \textsc{Quantum Espresso} (QE) package,\cite{QE} within the generalized gradient approximation (GGA) of Perdew and Wang (PW91) \cite{PW91-GGA} for the xc-functional.  For carbon atoms we used GGA-based ultra-soft pseudopotentials,\cite{pseudopotentials} and found the energy spectrum to be converged with a $30$~Ry plane-wave cutoff.  

To describe the molecule--surface interaction, we have used the quasiparticle $G_0W_0$ eigenvalues calculated with \textsc{vasp} along with the KS orbitals from \textsc{PWscf} and a somewhat reduced number of bands. For the determination of the screened interaction $W(\textbf{r},\textbf{r}',\omega=0)$, which enters into the BSE-Fock kernel (\ref{Fockker}), we used $300$ molecular orbitals, i.e.\ $120$ occupied and 
$180$ unoccupied orbitals. In order to obtain an accurate molecular absorption spectrum up to 10~eV, where the three most intense 
fullerene bright excitons lie\cite{Fullerene-opt1}, it is sufficient to include transitions within the fullerene $\pi-\pi^*$ complex in the 4-point polarizability. To solve the BSE (\ref{mateqforBSE}), we used a damping of $\eta = 50$~meV and a set of 14 occupied and 14 unoccupied states, i.e.\ $\{$HOMO$-13,$ HOMO$-12, \ldots,$ HOMO$,$ LUMO$, \ldots,$ LUMO$+12,$ LUMO$+13\}$.

To study a single isolated fullerene molecule, we must exclude the effect on its polarizability due to the interaction with the surrounding molecules in the lattice. This is 
accomplished in (\ref{indc}) by solving the BSE using a truncated Coulomb interaction \cite{RadialCutoff}
\begin{equation} 
V_C({\bf r}-{\bf r}')=\frac{\Theta\left(|{\bf r}-{\bf r}'|-R_C\right)}{|{\bf r}-{\bf r}'|},
\label{RadialCutoff}
\end{equation}
where $\Theta$ is the Heaviside step function, and $R_C$ is the range of the Coulomb interactions, i.e.\ the radial cutoff. Since the lattice constant $L \approx 24.18$~\AA\ is more than twice the range of the fullerene molecule's density, using a radial cutoff of $R_C=L/2$ ensures that the charge fluctuations created within the molecule produce a field throughout the whole molecule, but do not produce any field within the surrounding molecules. The definition (\ref{RadialCutoff}) is very useful because the Coulomb interaction remains translationally invariant. 

\section{Results and Discussion}
\label{Results}

\begin{figure}
\centering
\includegraphics[width=0.9\columnwidth]{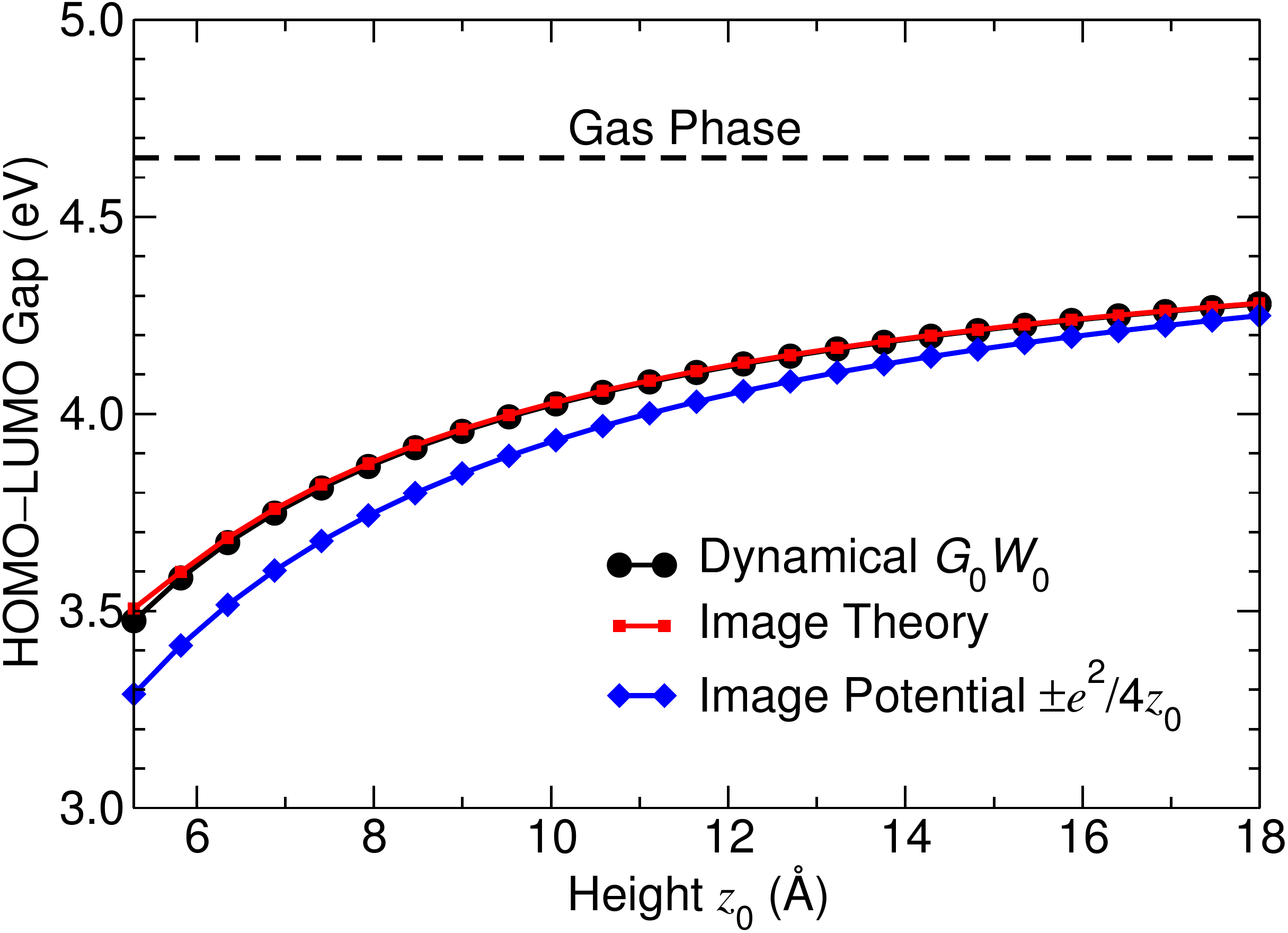}
\caption{(Color online) Calculated HOMO--LUMO gap of fullerene in eV as a function of the height $z_0$ in \AA\ above a metal surface using (black dots) the full dynamical $G_0W_0$ correction of Eqs.~(\ref{graingx},\ref{indcorr}), (red squares) image theory of Eq.~(\ref{imagecorr}), (blue diamonds) a HOMO and LUMO corrected by the image potential ($\pm\nicefrac{e^2}{4z_0}$). The gas phase result (black dashed line) is provided for comparison.}
\label{Fig2}
\end{figure}

We begin out analysis of the molecule--surface interaction by considering the quasiparticle $G_0W_0$ energy gap of fullerene near a gold or silver surface.  In Fig.~\ref{Fig2} we plot the fullerene HOMO--LUMO gap obtained using the dynamical $G_0W_0$ corrected by 
(\ref{graingx},\ref{indcorr}), as a function of the molecule's height $z_0$ above a jellium surface. We model the gold or silver metal surface using an electronic density parameter, i.e.\ the Wigner-Seitz radius, of $r_s\approx3.0a_0$\cite{AshcroftMermin}.

The quasiparticle HOMO--LUMO gap for fullerene in gas phase is 4.65~eV, somewhat lower ($\sim9\%$) than the experimental value of 5.1~eV\cite{HOMO--LUMO1,HOMO--LUMO2,HOMO--LUMO3}. We also compare with the HOMO--LUMO gap corrected using simple image theory which excludes dynamical effect in (\ref{graingx}) and (\ref{indcorr}). This model assumes the gas phase HOMO--LUMO gap is simply corrected by \cite{StevenLouie2006}: 
\begin{equation}
\frac{1}{2}\left\{
\Delta W^{LL}_{LL}(\omega=0)+\Delta W^{HH}_{HH}(\omega=0)\right\},
\label{imagecorr}
\end{equation} 
where $L=121$ is the LUMO and $H=120$ is the HOMO.  
We find this simple result agrees surprisingly well with the full dynamic $G_0W_0$ 
correction down to a height of $z_0\approx5.3$~\AA\ above the metal surface.  This means that the surface field does not create virtual 
transitions ($j\ne i$  in (\ref{graingx}) and (\ref{indcorr})).  Thus, the HOMO and LUMO behave as rigid charge distributions $|\psi_H|^2$ and $|\psi_L|^2$ which are screened by the static induced potential $\Delta W(\omega=0)$.  

The HOMO--LUMO gap is also shown in Fig.~\ref{Fig2} when HOMO and LUMO energies are corrected by an image potential of $+\nicefrac{e^2}{4z_0}$ and $-\nicefrac{e^2}{4z_0}$, respectively. We find even this very simple approach describes the HOMO--LUMO gap quite well for almost all heights considered. This means the fullerene HOMO and LUMO behave as positive and negative point charges at the center of the molecule down to $z_0 \gtrsim 5$~\AA.  

\begin{figure}
\centering
\includegraphics[width=\columnwidth]{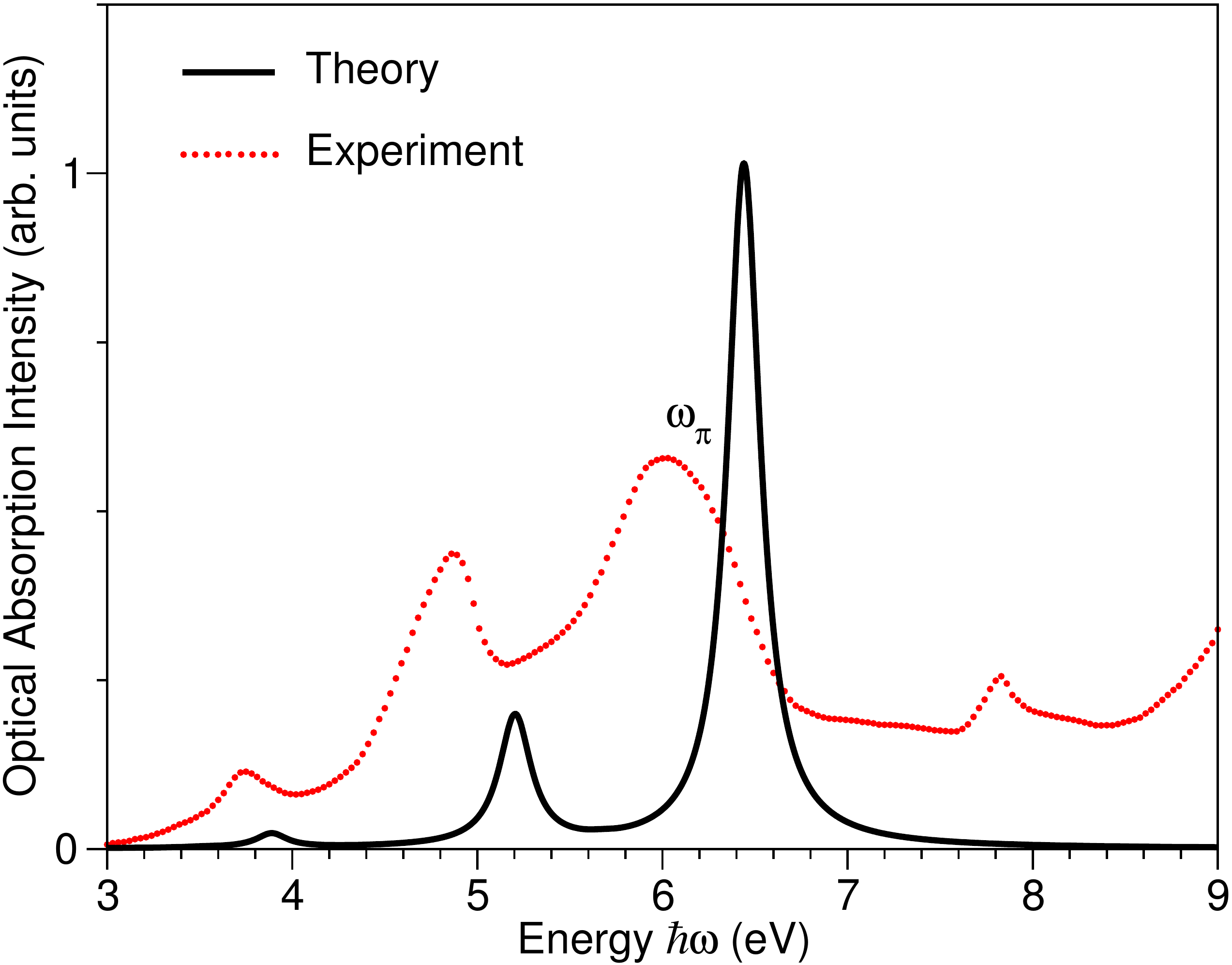}
\caption{(Color online) Optical absorption intensity for an isolated fullerene molecule as a function of the incident $x$-polarized light's energy $\hbar\omega$ in eV.  The calculated spectrum (black solid line) is compared with the measured spectrum from Ref.~\onlinecite{Fullerene-opt2}.}\label{Fig3}
\end{figure}

The optical absorption spectra shown in Fig.~\ref{Fig3} for an isolated fullerene molecule consists of three peaks.  These peaks correspond to the fullerene bright excitons observed experimentally at 3.77, 4.80 and 6.3~eV.\cite{Fullerene-opt1,Fullerene-opt2,Fullerene-opt3}. 
In our calculations the exciton energies are slightly blue shifted ($\sim 4\%$) to 3.9, 5.1 and 6.5~eV, respectively. 
The most intense peak at $\hbar\omega_{\pi} \approx 6.5$~eV corresponds to the fullerene $\pi$ plasmon resonance seen in EELS measurements \cite{Lucas1,Lucas2,FullereneEELS}. This is the optically active 
mode which is affected most by a metal surface.

\begin{figure}
\centering
\includegraphics[width=0.9\columnwidth]{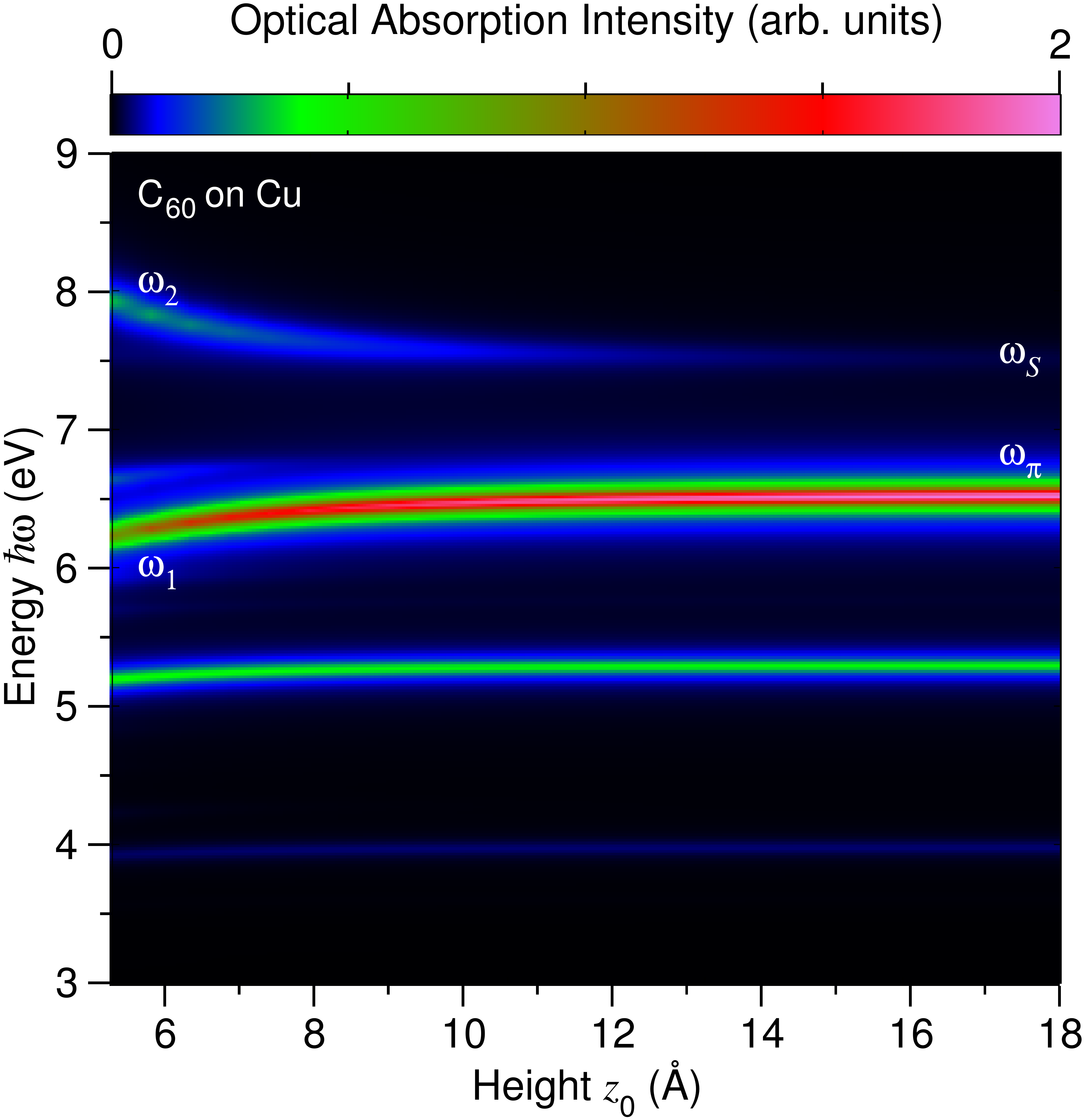}
\caption{(Color online) Fullerene optical absorption intensity as a function of the incident $x$-polarized light's energy $\hbar\omega$ in eV and the molecule's height $z_0$ in \AA\ above a copper surface ($r_s \approx 2.7a_0$).}
\label{Fig4} 
\end{figure}

In Fig.~\ref{Fig4} we show the fullerene optical absorption intensity as a function of the incident electromagnetic field frequency $\omega$ and molecule height $z_0$ for a copper surface. To model a copper surface the electronic density parameter, i.e.\ the Wigner-Seitz radius, is taken to be $r_s\approx2.7a_0$\cite{AshcroftMermin}. This corresponds to a surface plasmon energy of $\hbar\omega_S\approx7.5$~eV. 

Due to the high molecular symmetry of fullerene, we found that the absorption spectra of the isolated molecule does not depend on its orientation. When the molecule is close to the surface, the in-plane symmetry is unbroken.  For this reason, the absorption spectra does not depend on the direction of polarization in the $xy$-plane. However, the presence of the surface breaks the symmetry in the $z$-direction. We note that the surface weakly affects the isolated molecule's absorption spectra for $z$ polarized light. Here, we have chosen to apply incident light which is polarized in the $x$-direction.

When the molecule is far away from the surface ($z_0\approx18$~\AA), the absorption spectra corresponds to that of the isolated fullerene molecule shown in Fig.~\ref{Fig3}. 
As the molecule approaches the metal surface, the $\pi$ plasmon branch is weakened and pushed 
toward lower energies. At the same time, we find an extra branch appears in the optical absorption spectra.  When the molecule is far above the metal surface, this branch is horizontal 
and located exactly at the copper surface's plasmon energy $\hbar\omega_S\approx7.5$~eV. For smaller separations, this branch becomes more intense and 
disperses towards higher energies. These effects are a direct consequence of the interaction between the fullerene $\pi$ and copper surface plasmons.  
 
When the molecule is far above the metal surface, the interaction between molecular and surface charge oscillations is very weak. 
So if light of frequency $\omega_S$ excites currents in the molecule, even though the molecule responds only weakly to them, there 
still exists a weak Coulomb interaction with the surface.  It is this long-ranged interaction which channels the electromagnetic energy into a surface plasmon excitation. 
This is why the molecule is able to absorb light at the surface plasmon frequency $\omega_S$.

When the molecule is closer to the surface, the coupling is strengthened. The fullerene $\pi$ and copper surface plasmons hybridize and form 
coupled modes at frequencies $\omega_1$ and $\omega_2$. Modes $\omega_1$ and $\omega_2$ consist of 
$\pi$ plasma oscillations and ``localized'' plasma oscillations in the adjacent surface just below 
the molecule.  These modes oscillate out of phase and in phase, respectively, as sketched in Fig.~\ref{Fig8}(b). The molecular part of both modes has dipolar character.  This means the molecule absorbs  electromagnetic energy equally at both frequencies $\omega_1$ and $\omega_2$.   
\begin{figure}
\centering
\includegraphics[width=0.9\columnwidth]{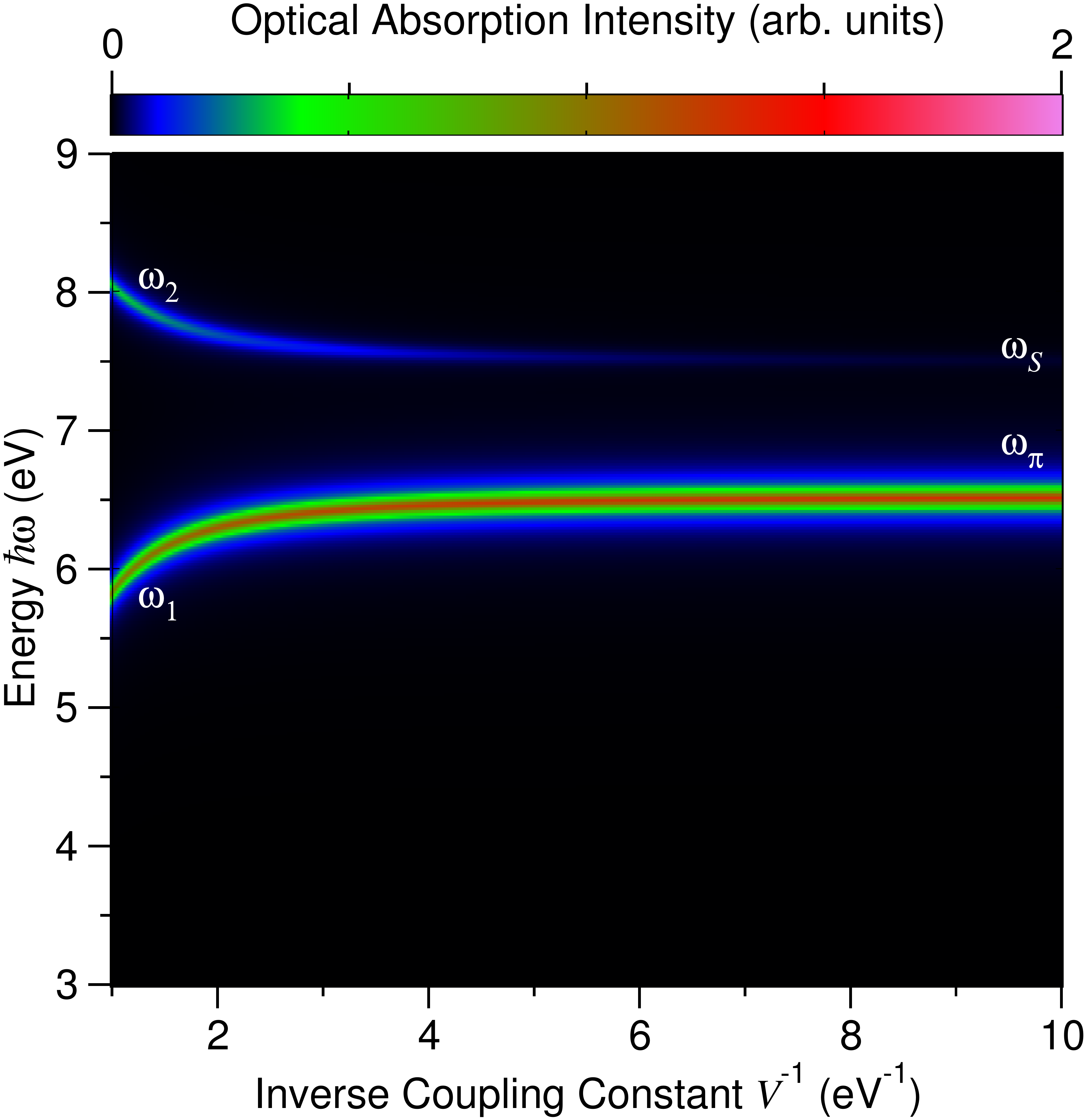}
\caption{(Color online) Fullerene optical absorption intensity as a function of the inverse coupling constant $V^{-1}$ in eV$^{-1}$ obtained from the point 
polarizable dipole model of Eqs.~(\ref{dipol1}--\ref{dipol3}). Dipole frequencies are chosen to be $\hbar\omega_\pi\approx6.50$~eV and $\hbar\omega_S\approx7.51$~eV, respectively. }
\label{Fig5} 
\end{figure}

If the interacting plasmas are spatially localized, then the optical absorption of the molecule should be well described by the point polarizable dipole model discussed in Sec.~\ref{pointdipole}.  In Fig.~\ref{Fig5} we show the molecular optical absorption intensities calculated with a point dipole model, as a function of the inverse coupling constant $V^{-1}$. The dipole frequencies are chosen 
to model a fullerene molecule ($\hbar\omega_M = \hbar\omega_\pi\approx6.50$~eV), and a copper surface ($\hbar\omega_S\approx7.51$~eV).  In this way we model the effect of a copper surface on the optical absorption of a fullerene molecule as a function of their coupling. 

As seen from Fig.~\ref{Fig5}, this model provides qualitative agreement with the absorption 
branches shown in Fig.~\ref{Fig4}.  This indicates that the optically active mode $\omega_1$ and the appearance of 
the new optically active mode $\omega_2$ are the result of an interaction between fullerene $\omega_\pi$ and ``localized'' copper surface $\omega_S$ plasmons.     

Below  the surface plasmon frequency $\omega_S$, the surface excitation spectra possesses a wide band of interband electron hole transitions 
\cite{Leo}. Since for copper the fullerene excitons lie below  $\omega_S$, they can interact with electron-hole excitations in the metal.   
However, in contrast to benzene bright excitons, which decay extraordinarily fast to electron-hole excitations \cite{Exciton,MoleculesSurfacePlasmonics},  we find the fullerene excitons do not interact with 
interband electron-hole excitations in the metal. This suggests that a more realistic description of the metallic surface would not influence the fullerene bright excitons significantly for the noncontact separations considered herein. 

\begin{figure}
\centering
\includegraphics[width=0.9\columnwidth]{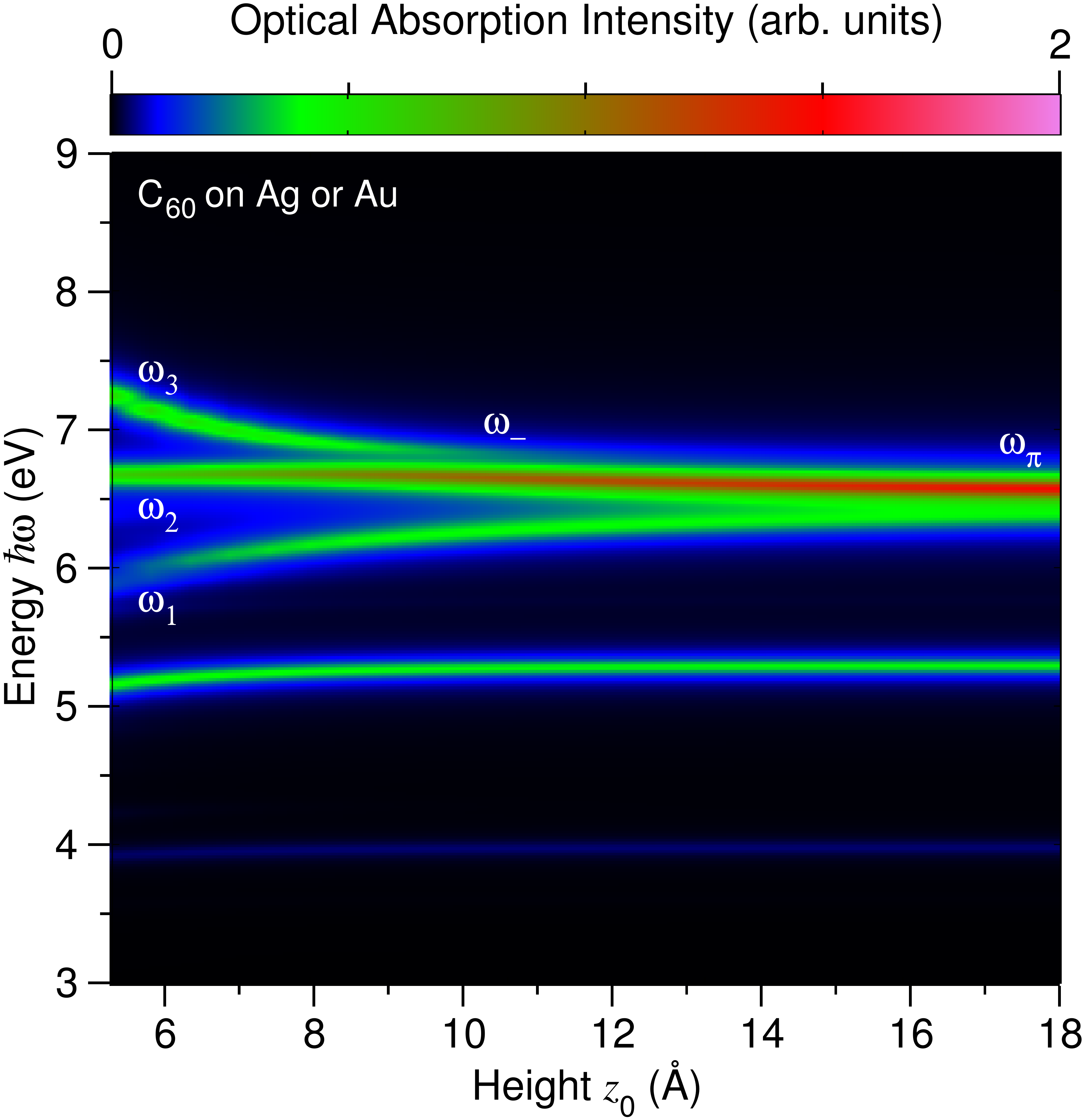}
\caption{(Color online) Fullerene optical absorption intensity as a function of the incident $x$-polarized light's energy $\hbar\omega$ in eV and the molecule's height $z_0$ in \AA\ above a silver or gold surface ($r_s\approx3.0a_0$).}
\label{Fig6}
\end{figure}

Figure~\ref{Fig6} shows the calculated fullerene optical absorption intensity as a function of the molecule's height above a gold or silver surface.  We again model the gold or silver metal surface using an electronic density parameter, i.e.\ the Wigner-Seitz radius, of $r_s\approx3.0a_0$\cite{AshcroftMermin}.  This corresponds to a surface plasmon energy of $\hbar\omega_S\approx6.41$~eV. In this case the fullerene $\pi$ plasmon is in resonance with the gold or silver surface plasmon, i.e.\ $\omega_\pi\approx\omega_S$.  

As the molecule approaches the surface, the two lower energy exciton branches at 4 and 5~eV are noticably bent towards lower energies. However, in this case, the absorption spectra for energies near that of the 
$\pi$ exciton shows an unusual behaviour. We find as the molecule approaches the surface one of the branches 
does not change in energy, while two new absorption branches separate from the $\pi$ exciton. In contrast to Fig.~\ref{Fig5}, 
where we have two absorption branches coming from hybridization between the fullerene $\pi$ and copper surface 
plasmon, here we have three branches.

To understand this phenomenon we next examine the symmetry of the
isolated fullerene $\pi$ plasmon modes. 
Because the linearly polarized electromagnetic field is a symmetric probe,
it is only able to excite symmetric modes which have dipolar 
character. This means antisymmetric (e.g.\ quadrupolar) modes are not excited. As a result, the optical absorption spectra does not provide a complete picture of the molecular excitation spectra.  To excite all types of modes requires an asymmetric probe. One way to do this is to examine the energy loss for an oscillating dipole placed close to the molecule. The formulation of the energy loss intensity for a point dipole placed in the vicinity of a molecule is given by Eqs.\ (62,63,71--73) in Ref.~\onlinecite{Exciton}. 

\begin{figure}
\centering
\includegraphics[width=\columnwidth]{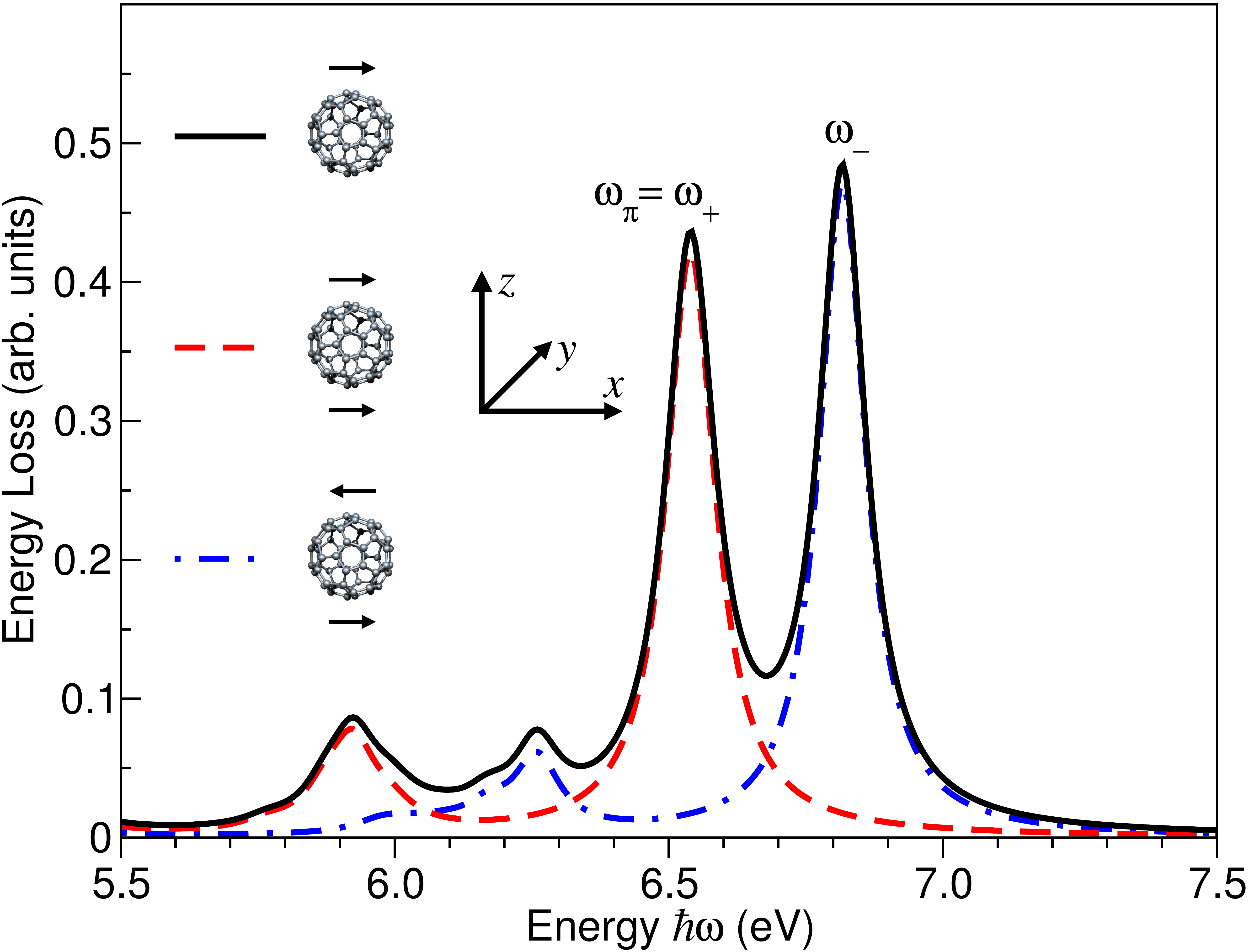}
\caption{(Color online) Dipole energy loss spectra as a function of the energy $\hbar\omega$ in eV of an $x$-polarized driving dipole placed at 6.35 \AA\ (black solid line) above, (red dashed line) in a symmetric configuration, and (blue dash-dotted line) in an antisymmetric configuration above and below the molecule's centroid.}
\label{Fig7}
\end{figure}

Figure~\ref{Fig7} shows the dipole energy loss of an  isolated fullerene molecule as a function of the driving frequency $\omega$. In all cases the dipole is $x$-polarized, i.e.\ ${\bf e}={\bf e}_x$ in (\ref{incidentemp}), and placed 6.35~\AA\ from the molecule's centroid. Fullerene has a clear plasmon mode at $\hbar\omega_+\approx6.50$~eV.  This corresponds to the $\pi$ plasmon frequency $\omega_\pi$ in the optical absorption spectra.  However, there is another strong plasmon peak at $\hbar\omega_-\approx6.8$~eV. 

In order to examine the parity of these modes we drive the molecule with two symmetrically placed dipoles, at 6.35~\AA\ above and below the fullerene's centroid. In one case the dipoles oscillate in phase, while in the other case 
they oscillate out of phase. When the dipoles oscillate in phase, they can only excite even parity molecular modes.  When the dipoles oscillate out of phase, they can only excite odd parity molecular modes. In this way we may clearly distinguish the character of the excited plasmon modes in the energy loss spectra.

Figure~\ref{Fig7} clearly shows when the dipoles oscillate in phase, the $\omega_+$ plasmon is excited, while  the $\omega_-$ plasmon is not. This is expected because the in-phase dipole oscillations mimic $x$-polarized light which only excites modes at $\omega_\pi\approx\omega_+$. However, when the molecule is driven by dipoles which oscillate out of phase, the $\omega_+$ plasmon is not excited, while the $\omega_-$ plasmon is excited. This means, in the isolated molecule, there exist two principal 
$\pi$ plasma oscillations; an $\omega_+$ mode with symmetric charge density oscillations (dipolar character) and an $\omega_-$ mode 
with antisymmetric charge density oscillations (quadrupolar character).  This is shown schematically in Fig.~\ref{Fig8}(a). Accordingly, because $\omega_+$ 
has dipolar character it is optically active, and because $\omega_-$ has quadrupole character it is optically inactive. 

\begin{figure}
\centering
\includegraphics[width=\columnwidth]{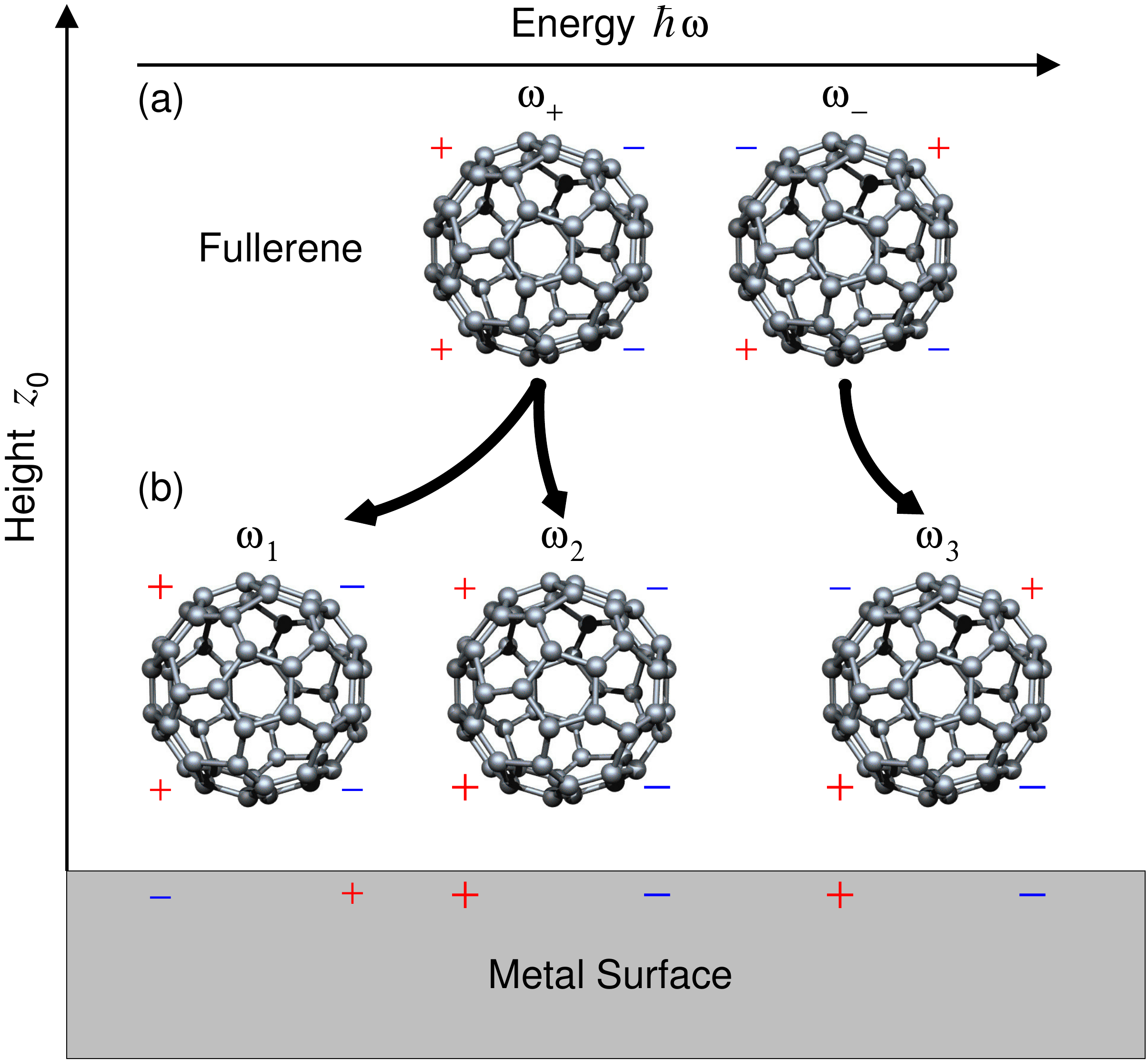}
\caption{(Color online) Induced charge density distributions of (a) an isolated fullerene molecule's  dipolar and quadrupolar modes $\omega_+$ and $\omega_-$, respectively, and (b) the fullerene/metal surface coupled modes $\omega_1$, $\omega_2$ and $\omega_3$.}
\label{Fig8}
\end{figure}

As fullerene gets even closer to the gold or silver surface, the molecular charge density oscillations start to interact with the surface plasmon. Because the $\omega_\pm$ plasmons are in resonance with the gold or silver surface plasmon $\omega_S$, the interaction is strong and destroys the $\omega_\pm$ plasmon's symmetry. This strong interaction with the metal surface plasmon splits the $\omega_\pm$ modes, destroys the pure quadrupole character of the $\omega_-$ mode, and  it begins to interact with the electromagnetic field, i.e.\ to be optically 
active. In this case two molecular modes $\omega_\pm$ and the surface plasmon $\omega_S$ form three coupled modes $\omega_1$, $\omega_2$ and 
$\omega_3$, all of which are optically active. 

The charge density distributions corresponding to modes $\omega_1$, $\omega_2$ and $\omega_3$ are sketched in Fig.~\ref{Fig8}(b). Mode $\omega_1$ is created from the molecular symmetric mode $\omega_+$, but oscillates out of phase with the surface charge density. Further, the induced charge is more localized in the upper molecular edge. Mode $\omega_2$ is created from the molecular symmetric 
mode $\omega_+$, and oscillates in phase with the surface charge density. In this case the induced charge is mostly localized in the molecule/surface 
interface. The third mode, $\omega_3$, is created from the molecular asymmetric mode, $\omega_-$, such that charge density in the molecule/metal interface 
oscillates in phase. 

Fig.~\ref{Fig6} shows that for larger separations ($z_0\approx 10.6$~\AA) the coupling is still too weak to split the $\omega_\pm$ modes. However, it is strong enough to destroy the $\omega_-$ quadrupolar character, making the $\omega_-$ mode optically active. This is manifested as a weak nondispersive branch slightly above the central peak at about 6.8~eV. Also, a small mode splitting 
$\Delta = \omega_--\omega_+\approx 0.3$~eV implies that there is a weak interaction between the charge density oscillations in the opposite halves of the molecule, as sketched in Fig.~\ref{Fig8}(a). 

Altogether, this suggests when fullerene is close to a copper surface, e.g.\ $z_0\approx 5.3$~\AA, the molecular optically active $\pi$ plasmon $\omega_\pi\approx\omega_+$ 
hybridizes with the metal surface plasmon $\omega_S$. This gives new coupled modes $\omega_1$ and $\omega_2$ which are both optically active. Moreover, when fullerene is close to a gold or silver metal surface, these modes are in resonance, i.e.\ $\omega_\pm\approx \omega_S$.  In this case the strong interaction with the surface plasmon destroys the purely quadrupolar character of the $\omega_-$ mode, and it begins to interact with the electromagnetic field. This leads to fullerene having three bright modes $\omega_1$, $\omega_2$ and $\omega_3$ in the $\omega_\pi$ region. This shows that the presence of a metal surface drives the optical activity of fullerene.

\section{Conclusion}
\label{Conclusions}
In this paper we have investigated how the interaction between a fullerene molecule and a coinage metal surface influences the optical activity of the molecule. We have shown that the interaction with the surface weakly affects the low energy fullerene bright excitons placed at 3.77 and 4.8~eV. However, the interaction with the fullerene $\pi$ plasmons is much more intense. 

In order to have a better understanding of this interaction, we first performed the point dipole energy loss calculation for the isolated molecule to determine the symmetry of the $\pi$ plasmons. We found that isolated fullerene supports two kinds of $\pi$ plasmons; a dipolar optically active mode at energy $\hbar\omega_+\approx6.5$~eV, and a quadrupolar optically inactive mode at a slightly blue shifted energy $\hbar\omega_-\approx6.8$~eV.

We have shown that when the molecule is close to a coinage metal surface $z_0\approx 5.3$~\AA, the dipolar plasmon $\omega_+$ hybridizes with the ``localized''  surface plasmon $\omega_S$. For a copper surface this produces two coupled modes  $\omega_1$ and $\omega_2$, which are both optically active. For a gold or silver surface the fullerene plasmons are in resonance with the surface plasmon, i.e.\ $\omega_S\approx\omega_\pm$.  In this case, the strong interaction with the surface plasmon destroys the purely quadrupolar character of $\omega_-$, and it also becomes an optically active mode $\omega_3$. 

Altogether, we conclude that the presence of a coinage metal surface enhances the optical activity of fullerene in the wide frequency interval around the intense $\pi$ plasmons.  These results have important applications in the areas of nanoplasmonic sensing of nearby molecules, and the engineering of fullerene-based photovoltaic materials.  

Our results clearly demonstrate that accurate computational screening of the molecule/substrate interaction is now possible within our BSE-$G_0W_0$ reformation.  This paves the way for the engineering of both the molecule and substrate in photovoltaic devices and nanoplasmonic sensors.  

\acknowledgments
V.D.\ is grateful to the Donostia International Physics Center (DIPC) and Pedro M. Echenique for their hospitality during various stages of this 
research. D.J.M.\ acknowledges funding through the Spanish ``Juan de la Cierva'' program (JCI-2010-08156), Spanish Grants (FIS2010-21282-C02-01) and (PIB2010US-00652), and ``Grupos Consolidados UPV/EHU del Gobierno Vasco'' (IT-578-13). The authors also thank I. Kup\v ci\' c for useful discussions.

%
\end{document}